\documentclass[aps,prd,twocolumn,superscriptaddress,groupedaddress,preprintnumbers,nofootinbib]{revtex4}

\usepackage{CJK}
\usepackage[english]{babel}
\usepackage{array}
\usepackage{tabulary}
\usepackage{amsfonts}
\usepackage{amssymb,amsmath}
\usepackage{cancel}
\usepackage{mathcomp}
\usepackage{graphicx}
\usepackage{color}
\usepackage{comment}
\usepackage{enumerate}
\usepackage{natbib}

\def\beq{\begin{equation}}
\def\eeq{\end{equation}}
\def\baq{\begin{align}}
\def\eaq{\end{align}}
\def\d{{\rm d}}

\begin{document}
\begin{CJK*}{}{}
\title{Small scale structures in coupled scalar field dark matter}

\preprint{}

\author{J.~Beyer$^a$ and C.~Wetterich$^a$ \\
{\it $^a$Institut f\"ur Theoretische Physik,   
Universit\"at Heidelberg,   
Philosophenweg 16, 69120 Heidelberg, Germany \\
}}

\begin{abstract}
\begin{center}
We investigate structure formation for ultralight scalar field dark matter coupled to quintessence, in particular the cosmon-bolon system. The linear power spectrum is computed by a numerical solution of the coupled field equations. We infer the substructure abundance within a Milky Way-like halo. Estimates of dark halo abundances from recent galaxy surveys imply a lower bound on the bolon mass of about $9 \times 10^{-22}$ eV. This seems to exclude a possible detection of scalar field dark matter through time variation in pulsar timing signals in the near future.
\end{center}
\end{abstract}

\maketitle

The cosmological standard model (or $\Lambda$CDM model) has provided a solid foundation for modern cosmology for a number of years by now. Still, the nature of two of its key components, dark matter and dark energy, remains unknown. So far both these components have eluded direct detection and can be seen only through gravitational effects.

In the $\Lambda$CDM scenario the dark sector consists of a pressureless fluid modeling cold dark matter and a cosmological constant making up dark energy. While the particle physics models generating a cold dark matter component are plentiful, the value of the cosmological constant $\Lambda$ is so tiny that is seems to contradict common expectations from quantum field theory. This is known as the \textit{cosmological constant problem}, which has prompted many investigations over the past years. A possible solution to this problem lies in dynamical theories of dark energy, most notably quintessence models, where a cosmological scalar field is used to describe dark energy \cite{Ratra:1987rm,Wetterich:1987fk,Wetterich:1987fm,Peebles:1987ek,Wetterich:1994bg,Copeland:2006wr}.

Despite its relative simplicity (and potential theoretical issues) the cosmological standard model has been very successful in explaining the vast majority of observed cosmological phenomena. Several predictions of the $\Lambda$CDM scenario for structure formation on small scales, however, have been claimed to be in conflict with increasingly precise cosmological observations. Most notable are probably the apparent predicted overabundance of dwarf galaxies in the Milky way galaxy (the \textit{missing satellite} problem \cite{Moore:1999nt,Klypin:1999uc}) and the cusp-like density profiles of halos which could be inconsistent with observed velocity dispersions both in galaxies (the \textit{cusp-core} problem \cite{Flores:1994gz,Navarro:1996gj}) and dwarf-galaxies (the \textit{too big to fail} problem \cite{BoylanKolchin:2011de,BoylanKolchin:2011dk,Springel:2008cc}). While some of these issues, in particular the missing satellite problem, might just be a result of our lack of understanding of the baryonic physics of galaxy formation \cite{Weinberg:2013aya,Genel:2014lma,Vogelsberger:2014dza,Vogelsberger:2014kha}, they may still be a hint towards possible modifications of dark sector physics. 

Amongst the many proposals that have emerged to solve these issues, warm dark matter (WDM) is probably the most popular one. If the dark matter particle is comparatively light (of the order of 1-4 keV) and is produced thermally in the early universe, it has a non-negligible velocity dispersion, thus suppressing the formation of structure on the relevant scales. This can solve some of the small scale problems of $\Lambda$CDM individually, as has been shown in several recent works \cite{Anderhalden:2012jc,Lovell:2011rd,Menci:2013ght,Papastergis:2011xe}. However, constructing a consistent model obeying all current observational constraints seems to be more difficult. In ref. \cite{Schneider:2013wwa} Schneider et al. argued that a WDM model consistent with all current observational constraints does not provide a significant improvement over cold dark matter predictions on small scales, at least not in the case of the simplest models of a single, thermally produced dark matter particle. One may therefore have to resort to more complicated scenarios of warm dark matter generation, or look for alternatives elsewhere.

Recently, we have proposed a unified picture of the dark sector, in which both dark energy and dark matter are modeled by scalar fields which couple through their common potential \cite{Beyer:2010mt}.
The mass of the scalar field responsible for dark matter was found to be somewhat larger than the inverse size of galaxies. In the present letter we show that the effects on structure formation are similar to WDM, thus establishing such a model as an interesting alternative explanation if small scale structures should indeed turn out to behave differently from the CDM expectations. Our model belongs to a general class of scalar field dark matter models which have been investigated in various incarnations \cite{Turner:1983he,Hu:2000ke,Matos:1998vk,Matos:2000ki,Matos:2000ss}. Besides its phenomenological interest it has the benefit of addressing the cosmological constant problem, as we will briefly discuss at the end. Furthermore, it provides for a natural explanation of a possible coupling between dark energy and dark matter (''coupled quintessence'' \cite{Wetterich:1994bg,Amendola:1999er}) which is often postulated somewhat ad hoc.


\subsubsection*{Class of models}
We consider two scalar fields $\varphi$ and $\chi$ with canonical kinetic terms and a common potential of the form
\beq
\label{commonPotential}
V(\varphi,\chi) = V_1(\varphi) + {\rm e}^{-2 \beta \varphi/M} V_2(\chi) \, ,
\eeq
with $M=2.44 \times 10^{18} $ GeV the reduced Planck mass.
We adopt the common name \textit{cosmon} for the quintessence field $\varphi$. The field $\chi$ is responsible for dark matter and dubbed \textit{bolon}, following earlier work \cite{Beyer:2010mt}.
The potential (\ref{commonPotential}) has been motivated by an investigation of possible consequences of approximate scale symmetry in higher dimensions \cite{Wetterich:2008bf,Wetterich:2009az,Wetterich:2010kd}. The dimensionless parameter $\beta$ will turn out to be the effective coupling between dark energy and dark matter.

The potential $V_1$ can in principle be any quintessence potential. For definiteness we use here an exponential potential
\beq
V_1 = M^4 {\rm e}^{-\alpha \varphi / M} \, .
\eeq
On the other hand, $V_2$ is restricted to an effectively quadratic shape at least in the late universe, where the field $\chi$ is supposed to act like dark matter \cite{Turner:1983he}. During the early stages of the cosmic evolution, $V_2$ can look very different indeed, and in fact a much steeper potential may be natural and desirable to ensure both insensitivity of the cosmic evolution on the precise initial conditions and stability of the adiabatic perturbation mode, which can be an issue in such coupled models \cite{Majerotto:2009np,EarlyScalings}. A shape very suitable for our purposes is the one proposed in ref. \cite{Matos:2000ss}
\beq
V_2(\chi) = c^2 M^4 \left( {\rm cosh} (\lambda \chi / M) -1 \right)\, ,
\eeq
which effectively matches an exponential to a quadratic potential and satisfies both criteria.

During the later stages of its evolution, when $V_2$ is effectively quadratic, $\chi$ follows a Klein-Gordon equation
\beq
\mathcal{D}_\mu \mathcal{D}^\mu \chi + m_\chi^2(\varphi) \chi = 0 \, ,
\label{kleinGordon}
\eeq
complicated by the fact that the mass is time-dependent as $\varphi$ increases with $t$,
\beq
m_\chi(\varphi) = m_0 {\rm e}^{- \beta \varphi / M} \, , \quad m_0 = cM\lambda \, .
\eeq
In a FLRW cosmological setting under rather generic assumptions, $\chi$ will oscillate quickly around its potential minimum, with its energy density scaling as 
\beq
\rho_\chi \propto a^{-3} {\rm e}^{-\beta \varphi / M}.
\eeq
The pressure $p_\chi$ on the other hand is highly oscillatory, but vanishes when averaged over a suitable timescale. At the background level such a coupled scalar field model results in a coupled quintessence cosmology \cite{Wetterich:1994bg,Amendola:1999er} for late enough times. 

For a wide range of initial conditions the value of $\chi$ at the later stages of the radiation dominated epoch depends only on the parameter $\lambda$ (and weakly on $\beta$), $\chi=\chi_0(\lambda,\beta)$, but not on the details of the initial conditions \cite{Beyer:2010mt}. (This differs from axion dark matter.) Typical values for $\chi_0$ are somewhat below the reduced Planck mass $M$. The bolon evolution towards the end of the radiation dominated epoch, as well as for the subsequent epochs, is therefore governed by three parameters, $m_0$, $\chi_0$ (or respectively $\lambda$) and $\beta$. One parameter has to be adapted in order to obtain the correct present matter density. For $\chi_0 \approx M$ one finds a typical present bolon mass of the order of the inverse galactic radius, while smaller $\chi_0$ yield somewhat larger masses \cite{Beyer:2010mt},
\beq
m_\chi^{-1} \approx 10 \left( \frac{\chi_0}{M} \right)^4 {\rm kpc} \, .
\eeq
Modifications of the CDM scenario on subgalactic scales therefore arise rather naturally in our setting.

\subsubsection*{Linear perturbations}
The evolution of linear perturbations around such a cosmological solution is rather intricate, as the oscillations present in the background interfere with oscillations in the perturbative quantities, making the averaging procedure a little tedious. A rigorous procedure results in an effective description which could have been guessed from the results of earlier works \cite{Hu:2000ke,Matos:2000ss}: The averaged perturbations behave like cold dark matter coupled to quintessence, but with a small (non-adiabatic) sound-speed present for large wavenumbers, which is given by
\beq
c_{s,\chi}^2 = \frac{k^2}{4 m_\chi^2(\varphi) a^2} \, .
\eeq
This sound-speed effectively suppresses the growth of modes with wavenumbers
\beq
\frac{k^2}{ a^2 H m_\chi(\varphi) } \gtrsim 1 \, ,
\eeq
which behave similar to photons. Here $H$ denotes the Hubble parameter.

For a given $k$ an effective growth sets in in the matter dominated epoch only once the scale factor $a$ exceeds $k/\sqrt{H m_\chi(\varphi)}$. The smallest wavenumber which is suppressed as compared to CDM can be roughly estimated by evaluating $k^2 = a^2 H m_\chi(\varphi)$ at the time when oscillations start (a similar scenario is described in ref. \cite{Matos:2000ss}). The corresponding Jeans wavenumber can be used to define a Jeans-mass, which is given by
\beq
M_J = \frac{4 \pi}{3} \left( \frac{m_0^2 {\rm e}^{-2 \beta \varphi^*/M} \rho_{r,0}}{3 M^2 (1-\Omega_\varphi^* - \Omega_\chi^*)} \right)^{-3/4} \rho_{\chi,0} \, ,
\eeq
where $\varphi^*$, $\Omega_\varphi^*$ and $\Omega_\chi^*$ are evaluated when oscillations start. (Sometimes a cutoff mass is defined by the wavenumber at which the linear power spectrum is suppressed by a factor of 2 compared to CDM. We choose to use a different notation for this cutoff mass, for which we find numerically $M_c \approx 3.3 \times M_J$, which is very similar to the result for WDM models discussed in \cite{Barkana:2001gr}.) 

We have computed the perturbation equations for the coupled cosmon-bolon system and solved them numerically, also including photons, neutrinos and baryons. Our code is a typical Boltzmann-code which uses manifestly gauge-invariant quantities and employs the same approximations as used in the CLASS-code \cite{Blas:2011rf,Lesgourgues:2011re} for photons, neutrinos and baryons, adapted to fit the gauge-invariant setting. The treatment of the scalar field sector needs a reliable map between the description by an oscillatory scalar field and the effective fluid description which is needed once the scalar oscillations become very rapid on the relevant time scales. We use the exact equations for the scalar field perturbations for both the bolon and the cosmon in the early universe and switch to the effective fluid description for the bolon given above for later times. The initial values for the fluid description are obtained by explicit numerical integration over one oscillation period in the field description. In case of the cosmon we have extended the approximation called \textit{radiation streaming approximation} in ref. \cite{Blas:2011rf} to the quintessence field for sufficiently late times. 

\begin{figure}[t]
	\centering
  \includegraphics[width=1.0\linewidth]{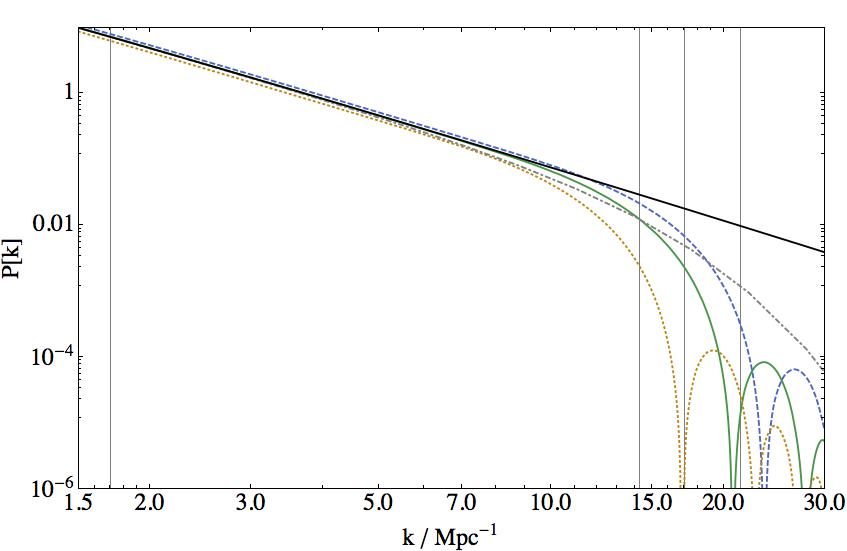}
	\caption{Power spectrum for a cosmon-bolon cosmology. We show results for different values of the coupling $\beta$. The solid green line stands for a bolon power spectrum with $\beta=0$ and $\lambda=65$, the dashed (blue) and dotted (orange) lines are obtained for the same $\lambda$ but with $\beta=0.05$ and $\beta=-0.1$ respectively. These curves represent (in the order described) present inverse bolon masses of $(4.9, 4.2, 6.4) \times 10^{-3}$ pc or $(7.7, 6.5, 10) \times 10^{20} {\rm eV}^{-1}$. For comparison, the solid black line represents a CDM power spectrum and the dotted-dashed line (gray) a WDM modification for a thermally produced WDM particle of mass $m_{\rm wdm} = 2.284$ keV. The three gridlines to the right correspond to the inverse length scale one would assign to a virialized spherical dark halo (with density contrast of $\delta_v = 200$) with radii of 8, 10 and 12 kpc via $k_r =  (\delta_v r^3)^{-1/3}$. The single gridline on the left corresponds to a radius of 100 kpc, roughly the size of the Milky way dark matter halo.}
	\label{fig:psCoupled}
\end{figure}

We have checked the accuracy of the effective fluid description by varying the time of transition from the field to the fluid description. We have confirmed that the matter-power spectrum in a $\Lambda$CDM cosmology obtained from our code agrees with the CLASS-result to excellent accuracy. More details can be found in ref. \cite{CB_LinPers}, where we present the analytical averaging procedure leading to the effective fluid description in detail and discuss the numerical treatment further. Our results for the power spectrum of adiabatic scalar fluctuations are shown in Fig. \ref{fig:psCoupled} for different values of $\beta$ and fixed $\lambda$. 

As is common in models of scalar field dark matter, the cutoff in the power spectrum is somewhat sharper than than in WDM-models. Notably, a positive coupling $\beta$ leads to a shift of the cutoff to larger wavenumbers. This effect originates from our adjustment the current bolon energy density $\rho_{\chi,0}$ to the fixed value $3 M^2 H_0^2 \Omega_{c,0} $, which effectively increases the bolon mass at the time when oscillations start for positive $\beta$ despite the ${\rm e}^{-\beta \varphi/M}$-dependence.

\subsubsection*{Small scale structures}
We are interested in how this kind of power-spectrum modification changes structure formation on small scales. A full investigation of structure formation would require high resolution runs of N-body codes. They would necessarily be based on effective (averaged) equations for the full non-linear perturbations. A more direct (and numerically much less involved) approach can be found by employing the extended Press-Schechter excursion set formalism (ePS) \cite{Press:1973iz,Bond:1990iw,Lacey:1993iv}. In this scenario hierarchical structure formation is modeled as a random walk of trajectories in density contrast space $\delta(S)$, where $S$ denotes the variance calculated from the linear power spectrum. The formation of a halo in this scenario is represented by the absorption of a trajectory by an absorbing barrier. The shape of the barrier has originally been assumed to be a constant, derived from the spherical collapse model. For some years now it has been known however, that a barrier modification motivated by an elliptical collapse is necessary to match the results of high-resolution N-body simulations \cite{Sheth:1999su,Sheth:2001dp}. Furthermore, in warm dark matter models, Barkana et al. showed that the barrier obtained from spherical collapse needs to be adjusted \cite{Barkana:2001gr}. The correct barrier exhibits a sharp upturn near the Jeans mass, a fitting formula is given by equation (7) in \cite{Benson:2012su}. Whether a similar upturn in the barrier is present in coupled scalar field models of dark matter as well remains an open question for now. We will merely illustrate the results such an effect might have on the predicted number of Milky Way subhalos.

The ePS-formalism does not directly yield predictions for numbers of subhalos within a given halo. It does however provide the conditional mass function $f(M_1,\omega(\delta_{sc,1},S) | M_2,\omega(\delta_{sc,2}),S)$, which describes the fraction of mass of a halo of mass $M_2$ at redshift $z_2$ corresponding to the barrier $\omega(\delta_{sc,2},S)$, which was contained in halos of mass $M_1<M_2$ at $z_1>z_2$ corresponding to the barrier $\omega(\delta_{sc,1},S)>\omega(\delta_{sc,2},S)$. As is common in ePS-analyses, we do not modify the power-spectrum when going to higher redshifts, but put all the time-dependence in the spherical collapse barrier instead, i.e.
\beq
\delta_{sc}(z) = \delta_{sc,0}/D(z) \, ,
\eeq
where $D(z)$ is the linear growth function. Furthermore we denote the elliptical barrier adjustment by the function 
\beq
\omega(\delta_{sc},S)=\sqrt{A} \delta_{sc} \left[ 1+b\left( \frac{S}{A \delta_{sc}^2} \right)^c \right] \, ,
\eeq
with $A=0.707$, $b=0.5$ and $c=0.6$.

Following \cite{Giocoli:2007gf}, we now calculate the current number of subhalos by a simple integration in barrier space, i.e.
\beq
\frac{\d n}{\d m} = \int_{\delta_0}^{\infty} \frac{M_2}{m} f(M_1,\omega(\delta_{sc,1})|M_2,\omega(\delta_{sc,2})) \d \delta_{sc,1} \, .
\eeq
When employing this procedure one loses the overall normalization, and we have to normalize the resulting number counts to N-body simulations. We used the CDM simulation in \cite{Lovell:2013ola} to adjust the $\Lambda$CDM curve, and employed the resulting normalization for all other models. 

The calculation of these first crossing rates needs to be done numerically for such complicated barriers, we used the recipe described in the appendix of \cite{Benson:2012su}. This approach treats the random walks as Markovian (i.e. uncorrelated), an assumption which is strictly speaking only true if one uses a sharp-k filter to obtain the variance function $S(M)$. However, as is well known, there is no unique way to assign a mass to a filtering radius for this choice of filter. As a result, huge uncertainties get introduced when one choses this filter, as we discuss in some detail in ref. \cite{CB_LinPers}. We therefore stick to a spatial tophat-filter, where this issue is not present. The price to pay is that the assumption of an uncorrelated random walk is incorrect in this case. One could calculate corrections resulting from the non-Markovian nature of the random walks \cite{Maggiore:2009rv,Maggiore:2009rw,Maggiore:2009rx,Farahi:2013fca,Musso:2013pha}, but based on the results of these investigations we expect them to be rather small compared to other effects neglected in this approach and do not do so here. Furthermore, the fitting of the elliptical barrier modification has been calculated in this way, too, and deviating from it would introduce additional uncertainties. 
\begin{figure}[t]
	\centering
  \includegraphics[width=1.0\linewidth]{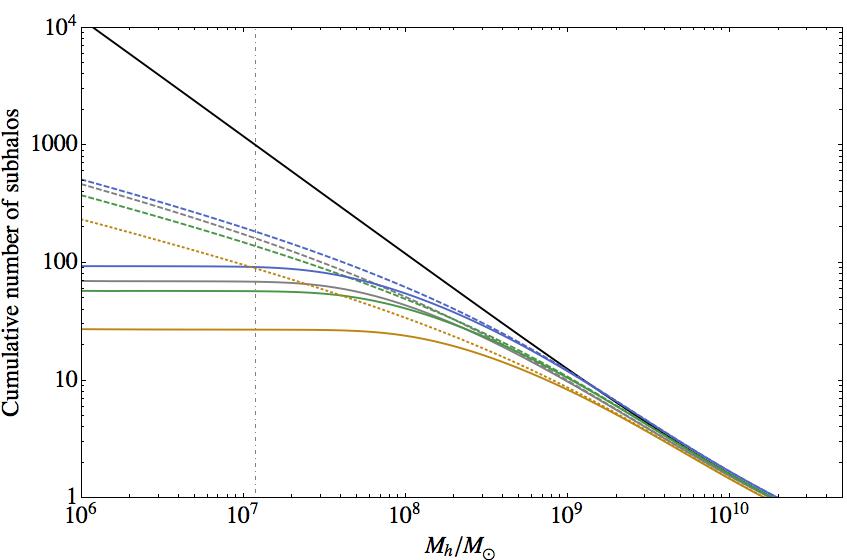}
	\caption{Cumulative number of Milky Way subhaloes as a function of halo mass $M_h$. The solid black line represents a CDM power spectrum and the gray lines a WDM modification for a thermally produced WDM particle of mass $m_{\rm wdm} = 2.284$ keV. The green lines stands for a bolon power spectrum with $\beta=0$ and $\lambda=65$, the blue and orange lines obtain for the same $\lambda$ but with $\beta=0.05$ and $\beta=-0.1$ respectively. For all WDM and bolon models, the dashed lines are results calculated without an upturn of the barrier near the Jeans mass, whereas we included such an upturn for the solid lines. As an additional orientation we added the dashed-dotted gridline at the WDM Jeans mass.}
	\label{fig:cumNums}
\end{figure}

The resulting cumulative number counts for several models are shown in Figure \ref{fig:cumNums} for a halo of mass $M_2 = 1.8 \times 10^{12} M_\odot$ at $z=0$. To gauge the accuracy of our calculations, one should compare the WDM-model with the results given in Figure 11 in ref. \cite{Lovell:2013ola}. Up to masses slightly above the Jeans mass our scenario seems to fit the N-body results rather well, but for smaller masses our curve stagnates whereas the N-body results continue to rise for a while longer. We seem to underestimate the asymptotic total number of subhalos by a factor of about 2. The reasons for this could be twofold:
First, structure formation is not strictly hierarchical, there are violent mergers and disruption processes present in N-body simulations, which can not be represented in the strictly hierarchical ePS-scenario. Such processes can generate halos even below the Jeans mass which can not form hierarchically. This might explain the underprediction of low mass halos we see in our approach.
Second, this is precisely the regime where spurious halos start to play a role in N-body codes, and uncertainties may arise in the identification of such halos.

One can use these results to put constraints on the allowed parameter range for our coupled cosmon-bolon model, simply by demanding that the number of subhalos should not fall below the number of dwarf galaxies estimated from observations. Estimates for this number range from 66 \cite{Lovell:2013ola} to several hundred \cite{Tollerud:2008ze}. For a bound on the bolon mass we choose the lower value of 66. We want to point out that galaxy formation for such small masses appears to be a highly stochastic process \cite{Strigari:2008ib}, potentially leaving a large number of halos void of stars, and the bounds we set here are therefore very conservative. 

\begin{figure}[t]
	\centering
  \includegraphics[width=0.92\linewidth]{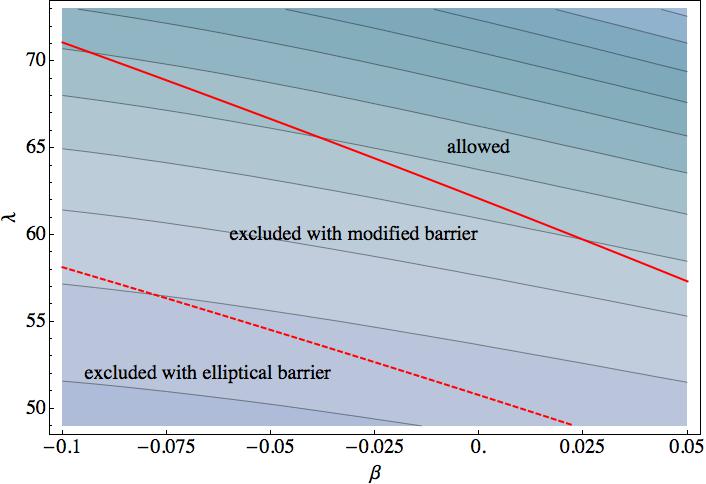}
  \raisebox{.5\height}{\includegraphics[width=0.06\linewidth]{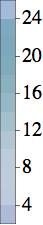}}
	\caption{Allowed parameter range for the cosmon-bolon model. The colored contours show different current bolon masses $m_{\chi}(t_0)$ in units of $10^{-22}$ eV. The solid red line displays the boundary of parameters which yield more than 66 subhalos in the Milky way if the modified barrier is used, the dashed red line shows the same exclusion curve for the standard elliptical barrier.}
	\label{fig:lbplot}
\end{figure}

The masses of the ultra-faint dwarf-galaxies appear to be universally around $ 10^7 M_\odot$, which is where we set our cut. At these masses, our method already underestimates the WDM N-body results already by a factor of roughly $1.5$ (we have checked this for all four models given in \cite{Lovell:2013ola}), so we artificially raise our obtained number counts by this factor when we use the modified spherical collapse barrier in order to remain extra cautious. The results can be seen in Figure \ref{fig:lbplot}.

Clearly larger couplings $\beta$ allow for smaller current bolon masses. Allowed values for the coupling strength are however constrained by CMB observations \cite{Pettorino:2012ts,Pettorino:2013oxa} to roughly $|\beta|\lesssim 0.1$. From this constraint we can derive an upper bound for current bolon mass, which we estimate by evaluating the boundaries presented in Figure \ref{fig:lbplot} for $\beta=0.05$:
\beq
m_{\chi}(t_0) \gtrsim 9.2 \, (4.1) \times 10^{-22} {\rm eV}
\eeq
for the modified (ellitpical) barrier. The typical scale at which we expect the formation of structures to be suppressed are however linked to the mass at $a^*$ and not today. Any mass can be related to a length scale via
\beq
m_{\chi}^{-1} \approx 0.64 \times \left( \frac{m_\chi}{10^{-23} {\rm eV}} \right)^{-1} {\rm pc} \, .
\eeq
Our bound lies within the typical range of ultra-light scalar field dark matter masses, but at the larger end, with important consequences for observational signatures.

\subsubsection*{Discussion}

In summary, we have presented an analysis of structure formation in the cosmon-bolon scenario of scalar field dark matter coupled to quintessence. We find that our scenario constitutes a valid alternative to standard cold dark matter. An interesting motivation of our model originates from higher dimensional theories of gravity. 
As was shown in a set of papers by one of the authors, higher dimensional dilatation symmetric scalar-tensor theories of gravity may provide an interesting solution to the cosmological constant problem \cite{Wetterich:2008bf,Wetterich:2009az,Wetterich:2010kd}.
For a scale invariant higher dimensional effective action one may consider the class of solutions to the field equations which allow for a reduction to a four dimensional theory. One finds that all stable quasistatic solutions of this type imply a vanishing effective four-dimensional cosmological constant. This "phase" in the space of solutions exists independently of the precise choice of the model parameters, reflecting a higher-dimensional mechanism of self-adjustment. 

Quantum fluctuations generically violate dilatation or scale symmetry due to the running of dimensionless couplings or mass ratios. For cosmological runaway solutions, however, a fixed point may be approached as the field expectation value which sets the scale of spontaneous scale symmetry breaking diverges for the asymptotic future $t \rightarrow \infty$. For a fixed point dilatation symmetry becomes exact. Thus the value of the four-dimensional effective potential has to vanish for $t \rightarrow \infty$, according to the phase-structure of higher dimensional scale invariant effective actions. In our present setting (Einstein frame) this limit corresponds to $\varphi \rightarrow \infty$. 

Spontaneous scale symmetry breaking implies the presence of an exactly massless Goldstone boson in the asymptotic limit $t \rightarrow \infty$ - the dilaton. Before this limit is reached, however, the effective potential does not yet vanish, inducing a small mass for the scalar field $\varphi$ which is associated with the cosmon responsible for dynamical dark energy. In our setting this is reflected by the exponential potential $V_1(\varphi)$. Depending on the geometry it may happen that a second long-range scalar field is present in the effective four-dimensional model, similar to moduli-fields in string theory. This second field is associated in our setting with the bolon field $\chi$. The vanishing of the effective potential in the asymptotic limit implies now for the combined potential $V(\varphi,\chi) \rightarrow 0$ for $\varphi \rightarrow \infty$, as realized by eq. (\ref{commonPotential}). It is precisely in this scenario where a coupling between the cosmon and scalar field dark matter arises naturally.

Our model has the potential to resolve possible shortcomings of small-scale structure formation in the 
$\Lambda$CDM model. In this context, recent findings showed that WDM is unlikely to be able to do so, at least in the case of a single thermally produced particle \cite{Schneider:2013wwa}. 
At first glance, our model predicts a scenario similar to WDM models. The onset of suppression of the power spectrum and the associated predicted number counts are almost the same for our model and WDM for a suitable choice of parameters. This similarity between scalar field dark matter and WDM at the linear level needs not extend to the non-linear evolution of perturbations. It is as of yet unclear how effective non-linear equations would look like for our model, and bounds derived for WDM from the Lyman-alpha forest \cite{Viel:2013fqw} might have to be recalculated for our scenario.  

If one looks at the internal structure of dark matter halos, the bolon model looks rather different from WDM. The scalar field oscillations are expected to translate to the gravitational potential in non-linear structures (similar to oscillatons or boson stars \cite{Lee:1995af,UrenaLopez:2001tw}), and such effects could in principle be detected. In a recent study Khmelnitsky and Rubakov investigated which mass range of a dark matter scalar field could lead to detection of such a signal through time variation in pulsar signals \cite{Khmelnitsky:2013lxt}. Our considerations of structure formation exclude this mass range by more than one order of magnitude (compare with Figure 1 in \cite{Khmelnitsky:2013lxt}). Even with conservative estimates, the mass of the scalar dark matter particle is probably too large to detect the scalar field oscillations in the forseeable future with pulsar timing signals.

\acknowledgments{The authors would like to thank Valery Rubakov for useful discussions. This work is supported by the grant ERC-AdG-290623.}

\bibliographystyle{unsrtnat}
\bibliography{letterbib}

\end{CJK*}
\end{document}